\documentclass[runningheads]{cl2emult}

\usepackage{graphicx} 
\usepackage{subeqnar} 
\usepackage{multicol} 
\usepackage{cropmark} 
\usepackage{lnp}      

\begin{document}

\title*{Effect of Excluded Volume and Anisotropy on \protect\newline
Granular Statistics: `Fermi Statistics' and \protect\newline Condensation}

\toctitle{Focusing of a Parallel Beam to Form a Point
\protect\newline in the Particle Deflection Plane}
%
%
\titlerunning{Fermi Statistics and Condensation}

\author{Daniel C. Hong}

\authorrunning{Daniel C. Hong}

\institute{Physics, Lewis Laboratory, Lehigh University, 
           Bethlehem, PA 18015, USA}

\maketitle              

\begin{abstract}
We explore the consequences of the excluded volume interaction of 
hard spheres at high densities and present a theory for excited granular
materials.  We first demonstrate that, in the presence of gravity,
the granular density crosses over from Boltzmann to Fermi statistics, 
when temperature is decreased in the weak excitation limit.  
Comparisons of numerical simulations with our predictions concerning the 
scaling behavior of temperature with agitation frequency, gravity and 
particle-diameter show satisfying agreement.
Next, within the framework of the Enskog theory of hard spheres, 
we interpret this crossover as a ``condensation'' of hard spheres from the 
dilute gas-state to a high density solid-like state.  
In the high density, low temperature limit Enskog theory fails because it
predicts densities larger than the closed packed
density below a certain temperature.
We show how to
extend the range of applicability of the Enskoq theory to arbitrarily low
temperatures by constructing a physical solution:
all particles that are situated in regions with
densities larger than a certain maximum density are assumed 
to be ``condensed''.
\end{abstract}

\section{Introduction}

This paper is a review of a recently proposed theory of granular 
dynamics~\cite{Haya,Hong}, which is based on the simple recognition that
granular materials are basically a collection of hard spheres that interact
with each other via a hard core potential~\cite{Jaeger}.
For this reason, many of the properties of excited granular materials 
may be understood from the atomistic view of molecular gases, in particular
from the point of view of kinetic theory ~\cite{Jenkins}.   
There are, however, several distinctions between
molecular gases and granular materials: First, the mean free path of 
the grains can be rather small -- less than the particle diameter -- 
even if the system is strongly forced.  Second, granular material is
made of macroscopic particles with finite diameter, and thus the material 
cannot be compressed indefinitely. When the mean free path vanishes, 
the density is the maximum, closed packed density.  
Third, gravity plays an important role in
the collective response of granular materials to external stimuli
because of the ordering of the particles according to their potential 
energy in the gravitational field.  

For example, one of the notable characteristics of 
excited granular materials in a confined system under gravity
is the appearance of a thin boundary layer near the surface that 
separates a fluidized region from a solid region.  
This effect is also
seen in shearing experiments~\cite{Hanes}, avalanches~\cite{Bou},
and grains subjected to weak excitations~\cite{Clement,Gallas}.  
In this limit, those grains in a solid region are effectively
frozen, and thus do not participate in dynamical, diffusive processes.  
Hence,
the conventional Boltzmann statistics, which is applicable in the limit of
strong excitation and rapid flow where all the particles are dynamically 
active, certainly needs modification.  Our first aim is to
show that, in the weak excitation limit, the statistics 
where such a boundary layer appears, is analogous to the Fermi
statistics.  We will in fact demonstrate that the density profile of
the grains is qualitatively well given by a Fermi function, from 
which we define the global temperature $T$ and develop a thermodynamic 
theory of configurational statistics for excited granular materials.
We present an explicit formula to relate the temperature $T$
to the external control parameters such as the frequency $\omega$ 
and amplitude $A$ of the excitation vibration, the diameter $D$ of
the grains, and the gravitational acceleration $g$.  
Next, we examine the microscopic
basis of the crossover from Boltzmann to Fermi statistics based on
Enskog theory for hard spheres~\cite{Enskoq}, and demonstrate how the 
crossover proceeds as grains condense from the bottom toward the surface.

\section{Configurational Statistics and Maximum Entropy Principle: 
Justification of a Thermodynamic Approach}

It is quite well known that variational methods are not useful 
in determining the properties of nonequilibrium systems.  In systems 
that show e.g.~periodic patterns, it is evident that one cannot perform
variations about a steady state situation. Since the system
being studied here is a dense, dissipative, nonequilibrium
granular system, we find it necessary to make some comments on this point.
If the mean free path of the grains is much less than a particle
diameter, each particle may be considered to be effectively confined
in a cage as also assumed for the free volume theory of dense 
liquids~\cite{Caram}.  In such a case, the basic granular state 
is not a gas, but a solid or a crystal \cite{Haya}. Thus an effective
thermodynamic theory based on the free energy argument may be more
appropriate than the kinetic theory.  
Our argument is that the dense state can be assumed as a ``steady state'' 
for which we compute the ``configurational statistics'' by means of
the usual variational method as the most probably state.

To be more specific, consider the excitation of disordered 
granular materials confined in a box with vibrations
of the bottom plate.  The vibrations
will inject energy into the system which cause the ``ground
state'' to become unstable. A new, excited state will emerge with
an expanded volume.  The time average of such configurations 
which have undergone structural distortions, may be deviating 
weakly from the ground state so that the use of an effective 
thermodynamic theory based on the variational principle could be 
justified.

Such a thermodynamic approach may be further
justified by the following two recent experiments in both the 
weakly and the strongly excited regimes:

\begin{itemize}
\item {\em Weakly or moderately excited regime}:
Cl\'ement and Rajchenbach(CR) ~\cite{Clement}
have performed an experiment with the vibrational strength,
i.e.~the dimensionless vibration acceleration 
$\Gamma = A\omega^2/g$, for a two dimensional vibrating bed,
using inclined side walls to suppress convection. 
CR have found that the ensemble-averaged density profile
as a function of height from the bottom layer obeys a universal function 
that is {\it independent} of the phase of oscillations of the 
vibrating plate. Namely, it is independent
of the kinetics imposed on the system.  One conceptually important point
here is that the reference point of the density profile is not the bottom
plate, but the bottom layer, which of course is fluidized.  

\item {\em Strongly excited regime}: Warr and Hansen (WH) \cite{Warr}
have performed an experiment on highly agitated, vertically vibrating 
beds of $\Gamma \approx 30-50$ using steel balls with a large 
coefficient of restitution.  They have found that the collective 
behavior of this vibrated granular medium
{\it in a  stationary nonequilibrium state}
exhibits strong similarities to those of an atomistic fluid
in  {\it thermal equilibrium} at the corresponding particle packing
fraction, in particular, concerning the two-point correlation 
function \cite{Warr}. 
\end{itemize}

The results of both experiments indicate that for both
moderately and strongly excited systems, a
one-to-one correspondence seems to exist between the 
{\it configurational} statistics of the {\it nonequilibrium} 
stationary state and the {\it equilibrium} thermal state.
In fact, this is not so surprising considering the fact that upon
vibration, the granular materials expand and consequently the volume 
of the system increases.
In turn, this increase corresponds to a rise in the potential
energy after the configurational average is appropriately taken.
Then the problem reduces to the packing problem, and the 
temperature-like variable, $T$, may be associated to the vibrating bed.
The existence of distinctive configurational statistics
in the density profile of CR (and also in WH in a special case)
appears to be fairly convincing evidence that kinetic aspects
of the excited granular materials
may be  separated out from the statistical configurations.
These observations are the basis of
the thermodynamic theory proposed in \cite{Haya}.  

\section{Fermi Statistics of Weakly Excited Granular Materials}

We first view the system of granular particles as a mixture of holes and
particles as in the lattice gas or the diffusing void model~\cite{Caram},
which is the simplest version of the free volume
theory~\cite{Hill}.   We now assign virtual lattice points by
dividing the vibrating bed of width $L$ and the height $\mu D$,
with $D$ typically the diameter of a grain and $\mu$ the number 
of layers, into cells of size $D \times D$.  Each row, $i$, is
then associated with the potential energy
$\epsilon_i=mgz_i$ with $z_i=(i-1/2)D$
and $m$ the mass of the grain.  Note that the degeneracy, $\Omega$, of
the each row is simply the number of available cells, i.e:
$\Omega=L/D$.  For a weakly excited system with $\Gamma \simeq 1$,
the most probable configuration should be determined by
the state that maximizes the entropy in the micro-canonical 
ensemble approach.

Taking into account the excluded volume effect which
plays a similar role for dense granular systems as the Pauli principle 
in Fermi statistics, we derive the entropy $S$, defined as the total 
number of ways of distributing $N$ particles into the system. 
Standard counting argument\ \cite{Huang} yields,
\begin{equation}
 S = \ln \, W 
   = \ln \left ( 
            \prod_i\frac{\Omega!}{N_i!(\Omega-N_i)!}
         \right ) ~,
\label {(1)}
\end{equation}
where $N_i$ is the number of particles occupying the $i-$th row.
Since gravity orders the grains with respect to their potential 
energy, a grain can be seen as a ``spinless Fermion'', where the 
height $z$ plays the role of the momentum variable (if one  
makes the connection to the electron gas).  
Maximizing $S$ with the constraints,  $\sum_i N_i =N $ and
$\sum_i N_i\epsilon_i = \langle U(T(\Gamma))\rangle$, the mean steady state
system energy, we find that the density profile, $\phi(z)$, which 
is the average number of occupied cells at a given energy level, is
given by the Fermi distribution \cite{Haya}:
\begin{equation}
\phi(z)=N_i/\Omega = 1/[1+\exp(\beta(z-\mu))] ,
\label{(2)}
\end{equation}
where $\beta=mgD/T$, the height is $z=z_i/D$ and analogous to the
Fermi energy, $\mu$, measured in units of $D$, is the number 
of layers in the low temperature limit, i.e.~in the system at rest.
Note that both $\mu$ and $T$ enter as Langrange multipliers introduced by
the two constraints, i.e, the conserved number of particles and the 
mean energy.  The global temperature T defined here is similar to
the compactivity introduced by Edwards and his collaborators in their
thermodynamic
theory of grains \cite{Edwards}, 
but is different from the kinetic temperature
defined through the kinetic energy \cite{Ther}.  We point 
out that the Fermi statistics is essentially the macroscopic
manifestation of the classical excluded volume effect and the anisotropy
which causes the ordering of potential energy by gravity. In the spirit
of the proposed analogy, the top surface of the granules at 
rest plays the role of the Fermi surface, and the grains in the 
thin boundary layer near the surface play the role of 
the excited electrons of a Fermi gas in metals.  

For strongly excited systems, the exclusion principle does not apply.
The Fermi analogy is valid when the zero temperature Fermi energy 
satisfies $\mu \gg n_l$, where $n_l$ is the number of fluidized layers.
Now, the energy $E_i$, injected into the system is of the order 
of $mA^2\omega^2/2$ and the potential energy $E_p$, needed
to fluidize the particles in the top $n_l$ layers is of the 
order of $mgn_lD$. Equating these two energies
we find a necessary condition for the fluidization of the top $n_l$
layers, namely: $n_l \sim \Gamma A/(2D)$.  Hence, we expect 
the Fermi statistics to be valid for $\Gamma \sim 1$,
if $\mu \gg A/D$.

\section{Relation between $T$ and $\Gamma$}

We now relate
the temperature $T$ to the external control parameters such as
$\Gamma$.  First, a thermal expansion.
We determine the energy per column $\bar u(T)
=\int_0^{\infty}\phi(z) mzg dz$, from which we can determine
the shift in the center of mass per particle;
$\bar h(T) = \bar u(T)/mg $, which is given by:
\begin{equation}
 \bar h(T) = h(0) \left [1+{\frac{\pi^2}{3}}
                        \left ( {\frac{T}{mgD\mu}} \right )^2
                  \right]
+\cdots
\label{(3)}
\end{equation}

\noindent with $h(0)=D\mu^2/2$.
Second, a kinetic expansion.  We make an observation that
for a weakly excited granular system, most excitations occur near the
Fermi surface, and thus the volume expansion may
be effectively well
represented by the maximum height, $H_0(\Gamma)$ of a  single
particle bouncing on the vibrating plate assuming that
the Fermi surface is in contact with the vibrating plate.
The kinetic expansion,
$H_0(\Gamma)$, is then determined by the maximum of $\Delta(t)$ in
the following equation that describes the
trajectory of a single ball on a vibrating
plane with 
intensity $\Gamma=\omega^2/g$ with A the amplitude, $\omega$ the 
frequency of the vibrating plate, and g the gravitational constant:

\begin{equation}
 \Delta (t) = \Gamma(\sin(t_0)-\sin(t)) + \Gamma \cos(t_0)(t-t_0) -
{\frac{1}{2}}(t-t_0)^2
\label{(4)}
\end{equation}
in units of $g=\omega=1$,where $t_0=\sin^{-1}(1/\Gamma)$.  Note that since
$\Delta$ is the relative distance between the ball and the plate, it
cannot be negative.  More precisely, the particle is launched from the plate
at $t=t_o$ by inertia, and then makes a free flight motion 
until it strikes the plate.  It then stays on the plate until
$t=t_o+2\pi$, when it is launched again and  repeats the same motion.  Hence,
$\Delta$ is a periodic function with period $2\pi$.
Since the Fermi distribution near $T=0$ can be approximated by a piecewise
linear function and $H_0(\Gamma)$ is thought to be the edge of the
function,  we expect
$H_0(\Gamma) \approx \Delta h/2
=(\bar h(T) - h(0))/2$.  By equating the thermal expansion,(\ref{(3)}),
 to the kinetic
expansion,
$H_0(\Gamma)g/\omega^2$, in physical units, we now complete our thermodynamic
formulation by presenting the explicit relation between $T$ and $\Gamma$ 
\cite{Haya}:

\begin{equation}
T = {\frac{mg}{\pi}} \left ( 3D \frac{ g H_0(\Gamma)}{\omega^2}
                      \right )^{1/2}.
\label{(5)}
\end{equation}
In MD simulation, one may measure the maximum height, $\bar h_o(\Gamma)$,
of a single ball on a vibrating plate and replace
$gH_o(\Gamma)/\omega^2$ with
$2\bar h_o(\Gamma)/\alpha$ with $\alpha$ an adjustable 
parameter~\cite{Quinn}.

Now we compare our theoretical prediction with an experimental result
of Cl\'ement and Rajchenbach \cite{Clement}.
Figure 1  shows the fitting of the
experimental density profile for $\Gamma=4$ of CR
by the scaled Fermi distribution, $\phi(z)=\rho(z)/\rho_c$,
with $\rho_c$ the closed packed density.  For hexagonal packing,
$\rho_c\approx 0.92$.
The fitting value of $T/mg$ is $2.0$\,mm,
while Eq.(\ref{(5)}) yields $T/mg \approx 2.6$\,mm.
The agreement between the two is fairly good in spite of
such a simple calculation.  This expression also agrees with the simulation
result~\cite{Gallas,Quinn}.

\begin{figure}
\centering
\includegraphics[width=.6\textwidth]{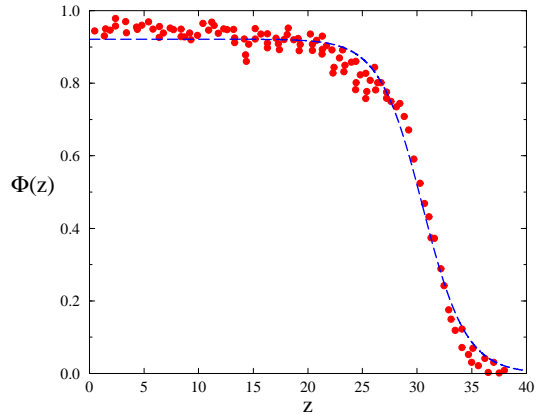}
\caption{ Density $\phi(z)$ as a function of the height z.
The symbols are the data by Cl\'ement and Rajchenbach and the 
dotted line is the Fermi distribution function, Eq.(2).
}
\label{eps1}
\end{figure}

Note that the
detailed expression of $H_0(\Gamma)$ depends on the manner by which the
grains are excited and we expect that our main scaling prediction
of Eq.(\ref{(5)}), namely
$T\propto g^{3/2}D^{1/2}/\omega$, will hold even for systems driven not
by sinusoidal waves.  Further, $T$ has a gap at $T=0$ because the time between
the launching and landing of the ball is always finite for $\Gamma >1$.
Next, it is well known that the
{\it specific heat} per particle, $C_v = d\bar u/dT$,
can be written as the fluctuations in the energy, namely 
$\langle (\Delta \bar u)^2 \rangle=\langle(\bar u(z)-\bar u)^2\rangle =
T^2C_v$ \cite{Nowak}. Hence, our theory makes a nontrivial prediction 
for the fluctuations in the center of mass:

\begin{equation}
\langle(\Delta z)^2\rangle = \langle (z(T)-\langle z \rangle)^2\rangle 
                           = \frac{\langle (\Delta h)^2 \rangle }{\mu_o^2}
                           = \frac{\pi^2}{3} \left (\frac{T}{mgD} \right )^3
                             \frac{D^2}{\mu_o^2} 
\label{(6)}
\end{equation}
while the center of mass is given by:
\begin{equation}
\Delta z(T) = z(T)-z(0) 
            = \frac{D\mu_o\pi^2}{6} \left (\frac{T}{mgD\mu_o} \right )^2 
\label{(7)}
\end{equation}
Note that the total expansion, $\Delta h(T)\equiv \mu_o\Delta z$ and its
fluctuations 
$\langle(\Delta h)^2\rangle/D^2$ = $\langle\mu_o(\Delta z)^2/D^2\rangle$ are
only a function of the dimensionless Fermi temperature $T_f=T/mgD$ as
expected.  Furthermore, note that (6) is an indirect confirmation that 
the specific heat is linear in $T$ as it is for the non-interacting Fermi gas.

\section{Test of Fermi Statistics by Molecular Dynamics Simulations}

In this section, we examine the configurational statistics of granular
materials in a vibrating bed, in particular the density profile, and its
fluctuations by MD simulations and compare the results with the predictions
made in the previous chapter.  The MD code was provided by the authors of
\cite{Lee} and
the details of the code can be found in the literature.  Simulations were
carried out in two dimensional boxes with vertical walls for $N$ particles each
with a diameter $D=0.2$\,cm and a mass $m=4\pi(D/2)^3/3$ with the degeneracy
$\Omega$ using $(N,\Omega)=(100,4),(200,4)$ and (200,8) with a sine wave
vibration. The dimensionless Fermi energy $\mu= N/\Omega$ is
the system height at rest.  For all cases, the inequality $\mu \gg A/D$ was
satisfied.  In Fig.~2 the temperature, obtained by the best fit of the
density profile to the Fermi function (dots), is plotted against $\Gamma$ 
and the values predicted by Eq.(5).  
Note the fairly good agreement between theory and simulations.
\begin{figure}
\centering
\includegraphics[width=.8\textwidth]{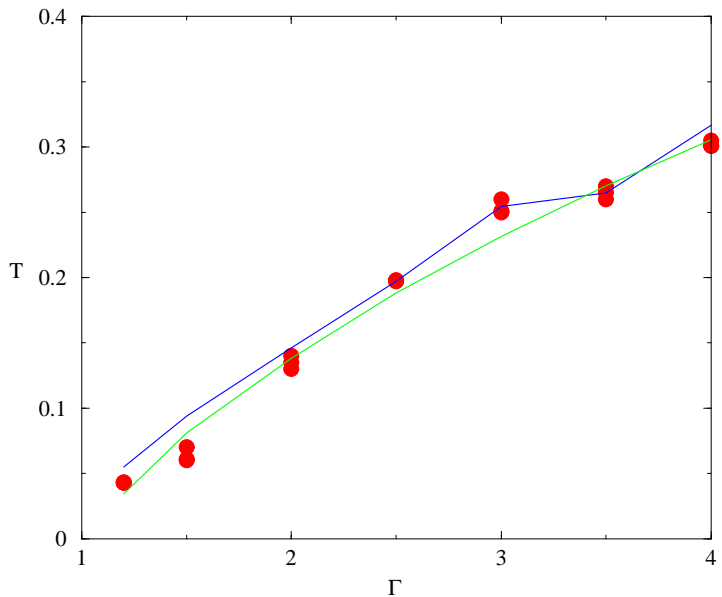}
\caption{ Comparison between the measured temperatures by
MD(dots) and the predicted ones.  The lower line was obtained by Eq.(5),
while the upper one was obtained by replacing $gH_o/\omega^2$ by 
$2\bar h_o(\Gamma)/\alpha$ with $\alpha\approx 12.8$.   $\bar h_o(\Gamma)$ 
is the maximum
jump height of a single ball on a vibrating plated obtained by MD.
} 
\label{fig2}
\end{figure}

We also studied the temperature scaling
against
the frequency, gravity and diameter to further check the validity of Eq.~(5).
The scaling laws as determined by the simulations are:\\
$$ T\approx \omega^{-m_1}$$
$$T\approx g^{m_2}$$
$$ T \approx D^{m_3}$$
with $m_1\approx 1.16$, $m_2\approx 0.48$, and $m_3\approx 0.53$. These values
are close to the predicted ones by Eq.~(5), i.e $m_1=1$, $m_2=0.5$, $m_3=0.5$.
For detailed comparisons, see the original paper~ \cite{Quinn}.

Finally, we have also checked the scaling of the center of mass, and 
the fluctuations against $T^2$ and $T^3$.  Since
the density profiles are well fitted by the Fermi function, we anticipate
that the center of mass and its fluctuations obey the scaling as shown in
Figs.~3 and 4.  Note that the increase in the center of mass is {\it second}
order in temperature T, which is contrary to the mean field prediction of 
the linear increase of volume in the compactivity X \cite{Edwards}.

\begin{figure}
\centering
\includegraphics[width=.8\textwidth]{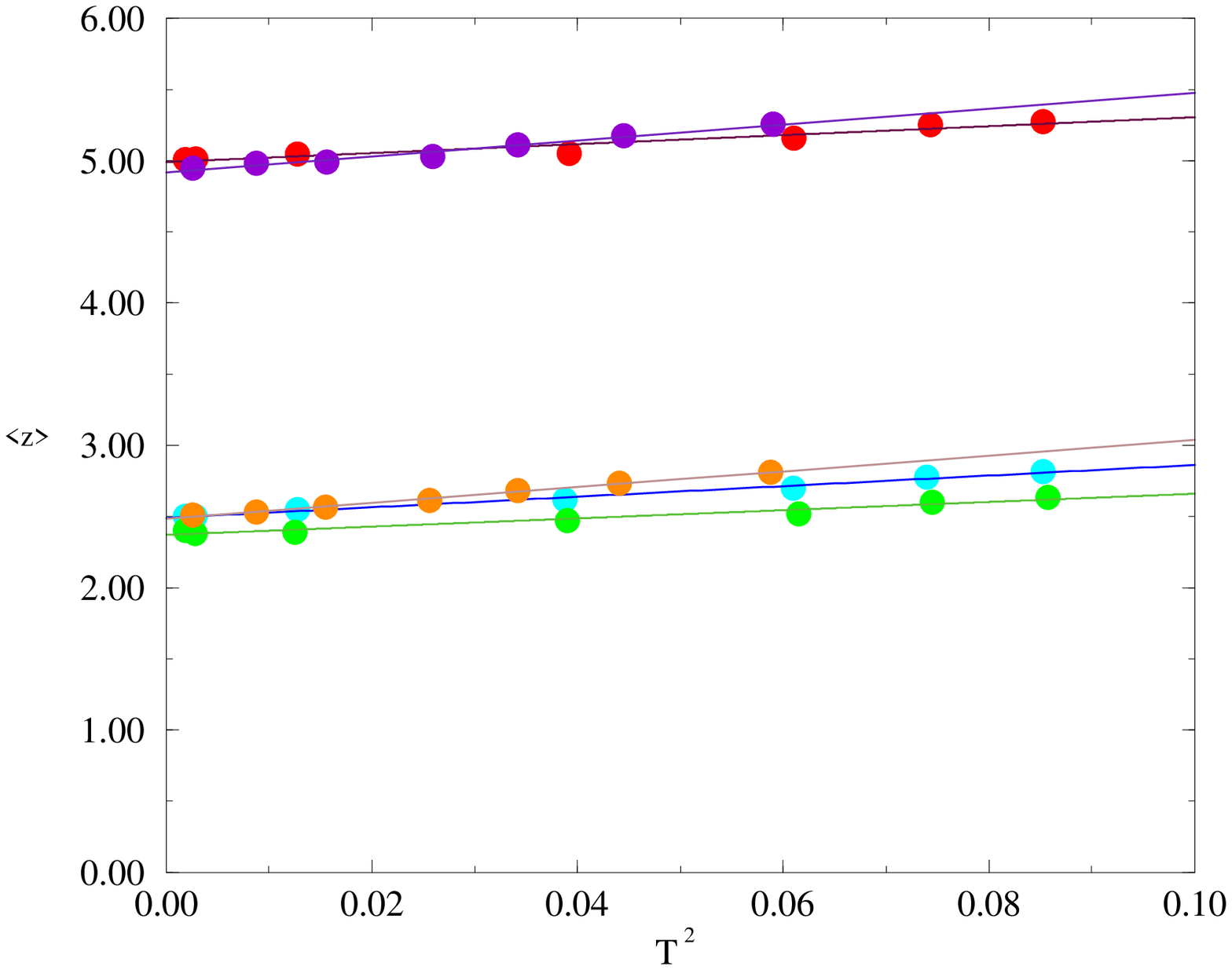}
\caption{ Center of mass as s function of temperature.  The upper ones
are data for sine wave and trianglular wave
excitations with $(N,\Omega)=(200,4)$ for different $\Gamma$'s.
The lower ones are the same with
$(N,\Omega)=(100,4)$. }
\label{fig4a}
\end{figure}

\begin{figure}
\centering
\includegraphics[width=.8\textwidth]{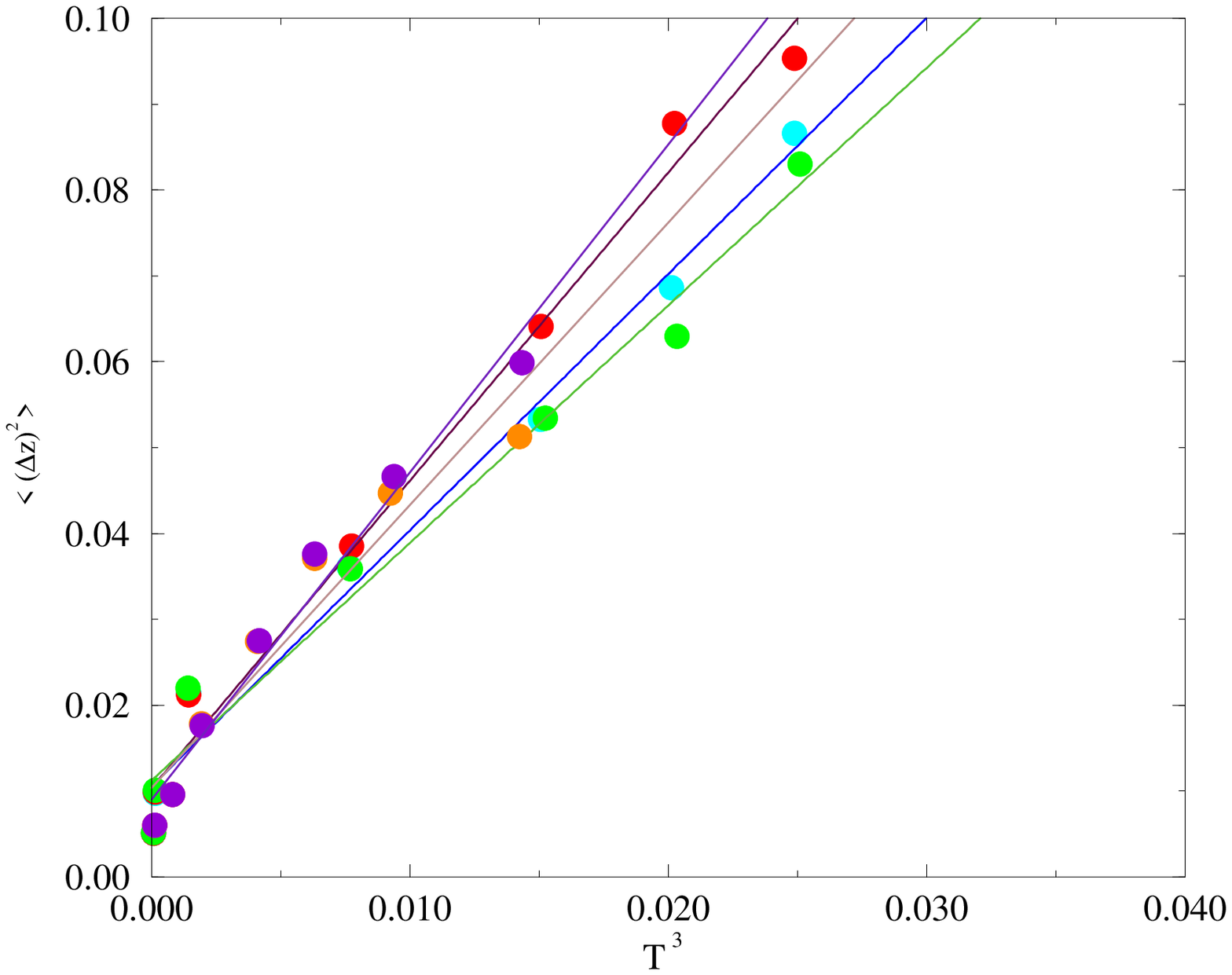}
\caption{ Fluctuations in the center of mass as a function of temperature.  
Symbols used are the same as in Fig.3.}
\label{fig4b}
\end{figure}

\section{Condensation of Hard Spheres Under Gravity}

Our next goal is to examine the microscopic basis of Fermi statistics 
based on the kinetic theory, in particular the Enskog equation 
for hard spheres of mass $m$ and diameter $D$, to explore
whether or not the kinetic theory can describe the 
cross over from Boltzmann to Fermi statistics and if so, 
under what conditions it occurs.  

Our particularly interesting discovery~\cite{Hong} is that
the prediction of the Enskog equation is only valid when
$\beta\mu \le \mu_o $, where $\mu$ is the dimensionless initial layer
thickness of the granules (or the Fermi energy),
$\beta=mgD/T$  with $T$ the temperature, and
the critical value, $\mu_o$, is determined to be
$\mu_o=21.756$ in two dimensions (2D) and $\mu_o=15.299$
in three dimensions (3D). 
(For maximum packing, $\mu_o=151.36$ in 3D
and $\mu_o=144.6155$ in 2D.)
When this inequality is violated, the
Enskog equation does not conserve the particles, and we present a scenario
of resolving this puzzle based on physical intuition, namely 
that the missing particles
condense from the bottom toward the surface~\cite{Hong}.
In this way, the hard sphere Enskog gas appears to contain
the essence of Fermi statistics. We briefly describe how the density
profile, $\phi$ as a function of dimensionless
variable $\zeta$, can be obtained from the
Enskog equation.  For details, see Ref. [2].
Note that $\phi$ introduced here is not the volume fraction $\nu$
but is scaled so that $\phi=1$ at closed packing.

In a free volume theory, particles are confined in a cage.  Hence, if we use
a simple cubic lattice as a basic lattice, the close packed volume fraction
$\rho_c = N/V = N/(D^2N) = 1/D^2$. 
If we define the dimensionless density
$\phi(z) =G(z)/\rho_c=D^2 G(z)$ or
$\phi(\zeta,\beta)=D^2 G(z)$
with $\zeta=z/D$,  
we then obtain the following exact dimensionless
equation of motion for $\phi(\zeta,\beta)$: 
\begin{equation}
{\frac{d\phi(\zeta)}{d\zeta}} + \beta\phi(\zeta) = \phi(\zeta) I_{\zeta}
(\zeta)
\label{(8)}
\end{equation}
where $\beta = mgD/T$ and
\begin{equation}
I_{\zeta}(\zeta) = {\frac{1}{2}}
\int_{0}^{2\pi}d\theta \cos\theta[\chi(\zeta-{\frac{1}{2}}\cos\theta)
\phi(\zeta-\cos\theta) - \chi(\zeta+{\frac{1}{2}}\cos\theta)\phi(\zeta
+\cos\theta)]
\label{(9)}
\end{equation}
For 3D, the corresponding equation
for the density $\phi(\zeta)=D^3 G(z)$ is given by Eq.(8) with:

\begin{equation}
 I_{\zeta}(\zeta)=
\pi\int^{\pi}_0 d\theta sin\theta \cos\theta [\chi(\zeta-\cos\theta/2)
\phi(\zeta-\cos\theta) - \chi(\zeta+\cos\theta/2)\phi(\zeta
+\cos\theta)] 
\label{(10)}
\end{equation}
Several forms for the equilibrium correlation function $\chi$ have been
proposed, but we use the following widely used forms:
For 2D, we use the form proposed by Ree and Hoover~\cite{Ree}:
$\chi(\phi) =(1-\alpha_1\phi + \alpha_2\phi^2)/(
(1-\alpha\phi)^2$ with $\alpha=0.489351\bullet\pi/2
\approx 0.76867, \alpha_1=0.196703\bullet\pi/2\approx 0.30898,
\alpha_2=0.006519\bullet\pi^2/4\approx 0.0168084$, while for
3D, we use the form suggested by Carnahan and Starling~\cite{Car}:
$\chi(\phi) = (1-\pi\phi/12)/(1-\pi\phi/6)^3$

Since the total number of particles, $N$, remains fixed,
the following normalization condition should be satisfied for both 2D and 3D.

\begin{equation}
\int_o^{\infty}d\zeta \phi(\zeta;\beta) = \mu 
\label{(11)}
\end{equation}

\noindent where $\mu \equiv N/\Omega_x$ (or $\mu\equiv
N/\Omega_x\Omega_y$ in 3D) is
the Fermi energy and $\Omega_x,\Omega_y$
are the degeneracies along the x and y axes.
We now perform the gradient expansion of (9) and (10) and retain only the
terms to first order in $d\chi/d\zeta$.  The 3D solutions for the first order
differential equation can be obtained easily, and are given by~\cite{Hong}:

\begin{equation}
-\beta(\zeta-\bar\mu) = ln\phi -1/(1-\alpha\phi)^2 + 2/(1-\alpha\phi)^3
\label{(12)}
\end{equation}

\begin{equation}
\beta\bar\mu = ln(\phi_o) -1/(1-\alpha\phi_o)^2 + 2/(1-\alpha\phi_o)^3
\label{(13)}
\end{equation}

\begin{equation}
 \beta\mu = \phi_o - \frac{2\phi_o}{1-\alpha\phi_o} + 
\frac{2\phi_o}{(1-\alpha\phi_o)^3}
\label{(14)}
\end{equation}

\noindent where $\alpha=\pi/6$.
For given values of $\beta$ and $\mu$, 
$\phi_o \equiv \phi(\zeta=0)$ will be determined
by Eq.(14) with the condition that $\phi_o$ cannot in anyway greater than
the closed packed density.  As we will see, we need care to proceed.
First, since the right hand side of (14) is an  
monotonically increasing function of $\phi_o $ and $\phi_o$ cannot
be greater than the closed packed density, which is 1 in our units,
$\beta\mu$ must have
an upper bound $\mu_o$; namely, $\mu_o=15.299$ in 3D,
($\mu_o=21.756$ in 2D), which is the value obtained by setting $\phi_o=1$
in the right hand side of (14).  Note that this upper limit
depends on the underlying basic lattice structure.  (For a hard sphere gas in
a continuum space, at the closed packed density, $\eta = \frac{4\pi}{3}
(\frac{D}{2})^3 = \frac{\pi}{6}\phi \approx 0.74$ 
in 3D and $\eta=\frac{\pi}{4}\phi= \pi/2\sqrt{3}\approx 0.907$ 
in 2D hexagonal packing, 
in which case, the upper limit $\mu_o= 151.36$ in 3D and 
$\mu_o=144.6155$ in 2D.)  Considering the
fact that both the temperature $T$ and
the Fermi energy $\mu$ are {\it arbitrary} control parameters, the existence
of such bounds is a puzzle: if $\beta\mu$ is less than $\mu_o$, then
the density profile given by Eq.(12) is well defined, but
if $\beta\mu$ is greater than $\mu_o$, then $\phi_o$ must be one at the bottom,
and the particle conservation appears to break down, namely

\begin{equation}
\int_o^1d\phi \zeta(\phi)=\int_0^{\infty}d\zeta\phi(\zeta) 
\equiv \mu_o/\beta < \mu 
\label{(15)}
\end{equation}

The central
question is: where does the rest of the particles go?  In order to gain
some insight into this question, consider first the case of point particles
under gravity, in which case the density profile is given by: $\rho(\zeta)=
\rho(0)
exp(-mg\zeta/T)$.  If we put more particles into the system, we simply
need to increase $\rho(0)$ because the point particles can be compressed 
indefinitely, and the profile
simply shifts to the right.  We now replace these point particles with
hard spheres, which cannot be compressed indefinitely and thus the maximum 
density at any point is the closed packed density.
Suppose we have a system of hard spheres at a certain temperature
T, where the density is closed packed
at the bottom layer and smoothly decreases toward the surface.  At this point,
if we add more hard spheres, say the amount of one layer,
to the system, how does the density profile 
modify?  Since the hard spheres cannot be compressed, our intuition tells us
that after the system reaches the equilibrium,
the density of the first(bottom) and second layer becomes closed packed, 
forming a rectangle, which is then
followed by the original Enskog profile.  If we
keep adding more particles, we obtain the
density profile that is the combination of the rectangle
(we term this the Fermi rectangle) beginning at
the bottom and 
the smooth original Enskog profile, which adjoins the Fermi rectangle at its
upper edge.
  The total number of
hard spheres in the Fermi rectangle should be the same as that of the
added particles.  One may obtain the same picture in a reverse way. 
Suppose we start from a high temperature
where all the particles are active.  We then slowly decrease the temperature 
to suppress the thermal motion.  There will be a temperature where the density
at the bottom is the closed packed density.  Now, let us lower the
temperature further.  What happens?  The next layer will become closed packed
and is thus effectively frozen.
As we keep lowering the temperature, the freezing of the
particles will occur from the bottom and the
frozen region will then spread out,
until at T=0 all the particles are frozen. 
Note that the frozen particles in the 
closed packed region behave like a solid with $\phi_o=1$.
Such an observation helps us to resolve the puzzle associated with the 
disappearance of particles.  The missing particles should
form the condensate in the Fermi rectangle spanning the bottom 
to the lower part of the fluidized layer.
We term the
surface that separates the frozen or a closed packed region with
$\phi=1$ from the fluidized region with $\phi <1$
the Fermi surface.  The location of the Fermi surface, $\zeta_F$,
is determined by
the number of the missing particles, namely, $\zeta_F=\mu-\mu_o/\beta$,
and is thus a function of temperature.
For nonzero $\zeta_F$, we must put the
missing particles below the Fermi surface and
shift the bottom layer from $\zeta=0$ to $\zeta_F$. 
From Eq.(15), we find that in 2D the
number of missing or condensed particles, $N_o(T)$, at $T <T_c$
 is $N_o(T)= \Omega_x(\mu - \mu_o/\beta) = N(1-T/T_c)$, 
where N($\equiv 
\Omega_x \mu$) is the total number of particles, 
and $T_c$ is the condensation temperature defined 
as the point where the particle conservation, Eq.(15), breaks down, namely

\begin{equation}
T_c= mgD\mu/\mu_o 
\label{(16)}
\end{equation}
Hence, the fraction of condensed particles is given by:
\begin{equation}
N_o(T)/N = 1-T/T_c
\end{equation}
How is this picture modified in the presence of dissipation?
With dissipation, solving the Enskog equation becomes a nontrivial task 
for three reasons: First, we do not yet know the precise functional 
form of the velocity distribution function for inelastic particles.
Experimentally observed profiles~\cite{Ola} definitely indicate that
they are not Gaussian.  Second, even if we somehow have empirical
formulas for the velocity distribution functions, carrying out the
integral with non-Gaussian profiles for the Enskog integro differential
operator becomes a non-trivial task.  Third, in the presence of dissipation,
the temperature profile is nonuniform~\cite{Grossman}, 
and thus one has to solve the
equation for the density profile, i.e., Eq.(8), along with the energy
equation~\cite{Haff}.  Analytic solution for this case is difficult to obtain.
However, if the dissipation is small,
then one might assume that the velocity distribution function is still 
Gaussian and the temperature remains uniform except near the
heat reservoir.  Under such assumption, we only need to solve the Eqs.(8)-(10)
with the corrected pressure term~\cite{Jenkins}, which is given by:

$$ P= \rho T[1 + \gamma\rho D^d \chi] $$
\noindent
where $\gamma = \frac{\pi}{4}(1+\epsilon)$ when $d=2$ and $\gamma = \frac{\pi}
{3}(1+\epsilon)$ when $d=3$.  Since the solution of the force balance 
equation, $dP/dz = -\rho g$ yields the same result, in the presence of
dissipation, we use the force balance equation instead of solving the
Enskog equation as was done in the elastic case~\cite{Hong}.
eDefining $\phi = \rho D^d$ as above and using the above 
forms of $P$ in the force balance equation yields the differential equation

\begin{equation}
\beta\phi(\zeta) + d\phi(\zeta)/d\zeta = -\gamma[\phi d\chi(\phi)/d\phi +  
       2\chi(\phi)d\phi(\zeta)/d\zeta] 
\label{(17)}
\end{equation}

Upon integration we find in the two dimensional case:

\begin{equation}
\beta(\zeta-\bar\mu)= -\log\phi + c_1\phi + c_2\log(1-\alpha\phi) + 
c_3/(1-\alpha\phi) + c_4/(1-\alpha\phi)^2 
\label{(18)}
\end{equation}

\noindent
where $$c_1 = - 2\gamma\alpha_2/\alpha^2,$$ $$c_2=\gamma(\alpha_1/\alpha^2 - 
2\alpha_2/\alpha^3)$$ $$c_3 = -c_2$$ $$c_4=\gamma(-1/\alpha + \alpha_1/
\alpha^2 -
 \alpha_2/\alpha^3)$$  
Then $\beta\bar\mu$ is the negative of the right hand side of (19)
with $\phi$ replaced with $\phi_0 \equiv \phi(\zeta=0)$.  
Integrating $\beta\zeta$ from zero to 
$\phi_0$ yields

\begin{equation}
\beta\mu = c_5 + c_6\phi_0 - c_2\phi_0^2 + c_7/(1-\alpha\phi_0) + 
c_8/(1-\alpha\phi_0)^2 
\label{(19)}
\end{equation}
\noindent
where $$c_5 = \gamma(-2\alpha_1\alpha + \alpha^2 + 3\alpha_2)/\alpha^4$$  
     
$$c_6= -c_2 + 1$$  $$c_7 = -\gamma(2\alpha^2-3\alpha_1\alpha + 4\alpha_2)/
\alpha^4$$              
$$c_8 = -c_4/\alpha$$
This result may be shown to be equivalent to the elastic case 
when $\epsilon = 1$.  
Substituting the numeric values for the constants $\alpha$, $\alpha_1$, 
$\alpha_2$ and evaluating at $\phi_0=1$ yields

\begin{equation}
\mu_0=1 + 10.3779(\epsilon+1) 
\label{(20)}
\end{equation}

In three dimensions we find

\begin{equation}
\beta(\zeta-\bar\mu)= -\log\phi - \frac{1}{4}\frac{\gamma}{\alpha
(1-\alpha\phi)^2} - \frac{1}{2}\frac{\gamma}{\alpha(1-\alpha\phi)^3} 
\label{(21)}
\end{equation}

\noindent
where $\beta\bar\mu$ is again
given by the negative of the right hand side of (22)
with $\phi_0$ replacing $\phi$.  Then we integrate $\beta\zeta$ between 
zero and $\phi_0$ to get

\begin{equation}
\beta\mu = \phi_0 + \frac{1}{2}\frac{\gamma\phi_0^2}{(1-\alpha\phi_0)^2} + 
\frac{1}{2}\frac{\gamma\phi_0^2}{(1-\alpha\phi_0)^3}
\label{(22)}
\end{equation}
\noindent
where $\alpha = \pi/6$.  When $\epsilon = 1$, this again
reduces to the profile for the elastic hard spheres (12).
We find upon evaluating $\beta\mu$ at $\phi_0=1$ that

\begin{equation}
\mu_0 = 1 + 7.14964(\epsilon + 1) 
\label{(23)}
\end{equation}

We note that in general, $\mu_0$ takes the form $1 + C(\epsilon + 1)$, and we 
recall that $T_c = mgD\mu/\mu_0$.  Hence we may write

\begin{equation}
T_c(\epsilon) = T_c(\epsilon=1) + \frac{mgD\mu C(1-\epsilon)}{(1+2C)(1+C(
\epsilon+1))}
\label{(24)}
\end{equation}

Letting $\delta = (1-\epsilon)\ll 1$, we may recast the above result as

\begin{equation}
T_c(\epsilon) = T_c(\epsilon=1) + 
\frac{mgD\mu C}{(1+2C)^2}\delta + \frac{mgD\mu C^2}{(1+2C)^3}\delta^2 + . . .
\label{(25)}
\end{equation}
In summary, we conclude that
the condensation phenomenon persists in the presence weak of 
dissipation.
Note that in the above analysis, we set $\phi_o=1$ as the upper bound
for a cage model.  For maximum packing, $\phi_o$ can exceed unity, as discussed
before, i.(in 3D, $\phi_o \approx 0.74\,\pi/6$ and $\phi_o=2/\sqrt{3}$.)
The clustering of particles near the bottom wall in
two dimensional experiments~\cite{Luding} may be a strong confirmation of this
scenario. 

\section{Conclusion}

In this review, we have explored simple consequences of excluded volume
interaction for dense granular materials, and shown 
that the granular statistics cross over 
from Boltzmann to Fermi statistics as the strength of the excitation is reduced.
We have also advanced a theory that such a crossover may be understood
as the condensation of hard spheres below a condensation 
temperature.  Since 
the Enskog equation breaks down at high densities, comparable to 
the closed packed density, 
we have constructed physical solutions beyond this regime by
forming the Fermi rectangle near the bottom.
Those particles in the Fermi rectangle are assumed to
be condensed.  This way,
we have extended the applicability of the Enskoq theory
to arbitrarily low temperatures.  

As demonstrated in Eq.(14), the Enskog solution based on
a virial expansion from the dilute to the dense case conserves particles
only when we allow the density profile greater than the maximum closed packed
density at the bottom, which is clearly unphysical.
This is due to the fact that the
equation of state, i.e.~the pressure as function of density, is not
known for all densities.  If one would use the correct equation for 
the pressure (which is not analytically known at this point, except for 
a numerically determined one~\cite{Luding2}) that must {\it diverge} at the
close packed density, it may be possible that the strange condensation
phenomenon associated with the Enskog equation may be absorbed by an 
advanced theory~\cite{Luding2}.  We point out, however, that 
even with this empirical pressure profile
the formation of Fermi rectangle is clearly visible.~\cite{Luding2}  
The density profile
above the Fermi rectangle was shown to be nicely fit by the Enskoq profile.  
Thus, our picture of condensation, which is an
alternative way to extend Enskog theory results up to the maximum 
possible density, seems to capture the essence of physics -- 
and we have shown analogies to the Fermi statistics.

For a polydisperse system, 
the upper limit for $\eta$ is usually larger than that of the 
monodisperse system.  Thus, we expect the freezing temperature to
be much lower for a polydisperse system.
Note, however, that as long as the upper limit for $\eta$ is smaller
than unity, Enskog theory for a gas predicts the condensation phenomenon,
in the sense that unphysical densities are predicted.  For the extreme
limit of an Appolonian packing~\cite{Appolo}, for which the close 
packed density can be unity, $\eta = \alpha\phi_o$ in (14) can be
arbitrarily close to unity and thus the right hand side of (14) can
be arbitrarily large.  In this case, the condensation phenomenon discussed
in this paper disappears.  

The original experiment done in Ref.~\cite{Clement} used a rectangular 
box with tilted side walls, but one may use a different shape such as a
parabola to mimic the density profile of the electron gas in 3D, which may 
help one to visualize the excitation spectrum near the Fermi surface.
Finally, we point out that while it is trivial to drive hard sphere 
grains by a thermal reservoir in MD simulations, it becomes a nontrivial 
task to do this experimentally.  The normal
way of exciting the granular system experimentally
is through vibration, yet it remains
an unresolved issue to determine the precise relation
between the vibration strength and the effective temperature of the 
reservoir.  Eq.(5), which is based on a single ball picture,
may be a first step in this direction.
The discovery of the Fermi statistics and the associated
condensation phenomena may provide an avenue to
apply the methods developed in equilibrium statistical mechanics
to the study of granular dynamics, notably
from the point of view of elementary excitations such as the Fermi liquid
theory in condensed matter physics.

\section*{Acknowledgments}
I wish to thank H. Hayakawa for collaboration and J. A. McLennan for
helpful discussions on the Enskog equation.  
I also wish to thank Paul Quinn for carrying out
MD simulations and Joseph Both for checking some of the algebra in section 
6, as well as S. Luding for discussions on the 
Appolonian packing and the puzzle associated with the condensation phenomenon.


\begin{thebibliography}{7}
\bibitem{Exc} This is similar to the case of
electron gas subjected to thermal excitation, in which case
only electrons near the Fermi surface are excited.  Note that while the
exclusion principle operates in momentum space for an electron gas, 
for grains, the corresponding exclusion principle operates in real space.
\end{thebibliography}
\end{document}